\documentclass[12pt]{article}
\usepackage{amssymb,amscd,amsmath,amsthm}

\usepackage{latexsym,amstext}
\usepackage{color}
 \usepackage{graphicx}
\begin{document}

\title{\bf Evolution of quantum observables: from non-commutativity to commutativity}

\author{ S. Fortin$^1$, M. Gadella$^2$, F. Holik$^3$ and M. Losada$^1$\thanks{All authors have contributed equally to this work}}

\maketitle

$^1$CONICET, Universidad de Buenos Aires, Buenos Aires, Argentina.

$^2$ Departamento de F\'isica Te\'orica, At\'omica y \'Optica and IMUVA,
Universidad de Valladolid, Paseo Bel\'en 7, 47011 Valladolid, Spain.

$^3$ Instituto de F\'isica La Plata,
Consejo Nacional de Investigaciones Cient\'ificas y T\'ecnicas,
La Plata, Buenos Aires, Argentina

\bigskip

\begin{abstract}

A fundamental aspect of the quantum-to-classical limit is the transition from a non-commutative algebra of observables to commutative one.  
However, this transition is not possible if we only consider unitary evolutions. 
One way to describe this transition is to consider the Gamow vectors, which introduce exponential decays in the evolution. 
In this paper, we give two mathematical models in which this transition happens in the infinite time limit.
In the first one, we consider operators acting on the space of the Gamow vectors, which represent quantum resonances.
In the second one, we use an algebraic formalism from scattering theory. We construct 
a non-commuting algebra which commutes in the infinite time limit.

\end{abstract}

\section{Introduction}

In previous papers we have discussed an alternative approach to the quantum-to-classical transition of physical systems based on the evolution of the algebra of observables. A non-commutative algebra of observables evolves into a commutative
one, when time tends to infinity.
However, this is not
possible if the time evolution is unitary. In order to achieve a transition from a non-commutative algebra to commutative one, we need to consider more general time evolutions.

This approach has been applied in the context of self-induced decoherence \cite{FV, Fortin2013}, enviroment induced decoherence
 \cite{Schlo, Fortin2010, LFH}, and for quantum maps \cite{LFGH}. Also, we have studied its formal aspects and its  algebraic properties \cite{LFGH, FHV}. 

 In \cite{LFGH}, we have
given an example based in the Gamow formalism for resonances. The
space of states is formed by the Gamow vectors representing
resonance states. The algebra of operators is given by the
linear mappings on the space of this resonance states. In order to get a
suitable non-unitary time evolution, it is necessary to isolate the
resonances from the environment, as described in \cite{LFGH}. This
is summarized in Section 2 of this work.  It is
important to recall that resonances represent quantum unstable
states \cite{BEU}, we shall use these two terms indistinctly.

In this paper, we give a further generalization of the non-unitary
evolution presented in \cite{LFGH}. This is done by appealing to a
formalism for quantum observables and states that was originally
developed so as to include van Hove states, originally conceived to be applied in statistical mechanics of systems far from equilibrium. In this
formalism, observables belong to a non-commutative algebra and
states are represented by positive functionals over the algebra of observables. Then, we discuss the scattering process given by a
Hamiltonian pair $\{H_0,H=H_0+V\}$, where the scattering is produced
by a potential $V$. This gives place to two isomorphic algebras
which can be put in connection with the \textit{in} and \textit{out} states of
the scattering process.  Although resonances may appear in this model
and Gamow states are properly defined, we do not rely in the
presence of resonances this time. Under suitable conditions, \textit{in
observables} and \textit{out observables} commute at the limit $t\longmapsto
-\infty$ and $t\longmapsto +\infty$, respectively.

This paper is organized as follows: In Section 2, we discuss a model
based in the construction of the Gamow spaces with Krein
pseudo-metrics and we show that in the limit $t\longmapsto\infty$ all
operators commute. In Section 3, we describe the construction of in
and out algebras and how this algebras become commutative at the
limits of large negative and positive times. Finally, we present our conclusions.

\section{Quantum resonances}

There are several definitions of
quantum resonances, not all of them equivalent \cite{FGR,BOH,GAD},
based either on physical or mathematical considerations. From the
physical point of view, one may assume that resonances come from
resonant scattering: an incident particle stays a long time on an
interaction region and then scapes. Thus, from the physical point of
view resonances can be detected \cite{BOH} either by long Wigner
times \cite{GKN}, by a sharp bump in the cross section, or a
sudden change in the phase shift of about $\pi/2$.

The resonance scattering requires of the existence of two dynamics,
a {\it free} dynamics governed by a Hamiltonian $H_0$ and a {\it
perturbed} dynamics governed by a total Hamiltonian $H=H_0+V$,
where $V$ is the potential giving by the perturbation which is
responsible for the resonance behaviour.

From the mathematical point of view, there are two definitions of
resonances, not always equivalent (although in many models they are
equivalent). One of them is based on the study of the poles appearing in the resolvent functins associated to the Hamiltonians, and the other is based on the S-matrix formalism.

For the former, we define the following pair of complex functions
for $\psi\in\mathcal H$, where $\mathcal H$ is the Hilbert space of
the system under consideration \cite{RSIV}:

\begin{equation}\label{1}
F_0(z):= \langle \psi|(H_0-z)|\psi\rangle\,, \qquad F(z):= \langle\psi|(H-z)|\psi\rangle\,.
\end{equation}
Then, assume that there is a dense subspace $\mathcal D$ such that for any $\psi\in\mathcal D$ these functions are
meromorphic having the positive semi-axis $\mathbb R^+$ as a branch cut and that they are analytically continuable through the cut. Then, if $F(z)$ has a pole in the analytic continuation at $z_R=E_R-i\Gamma/2$ and $F_0(z)$ is analytic at this point, we say that the Hamiltonian pair $\{H_0,H\}$ has a resonance pole at $z_R$. Its real part $E_R$ is the resonance energy and $\Gamma$ is the inverse of the mean life. Note that the complex conjugate $z_R^*=E_R+i\gamma/2$ of $z_R$ has the same property.

The second definition of resonance assumes that the $S$-matrix has an analytic continuation either as a function of the momentum $p$ or of the energy $E$ \cite{BOH, NUS}. In this case, which is the one we will consider in this paper, resonances are characterized by poles of the analytic continuation of $S(E)$ through the cut $\mathbb R^+$. These poles appear as complex conjugate pairs, $z_R=E_R-i\Gamma/2$ and $z_R^*=E_R+i\Gamma/2$, both representing the same resonance. In many relevant models, such as the Friedrichs model \cite{F,EX,GP}, poles of $S(E)$ and poles of $F(z)$ coincide. However, there are models for which poles of $S(E)$ does not correspond to bumps in the cross section and vice-versa \cite{FGR}.

Pure states are represented by unit vectors in the Hilbert space $\mathcal H$. Now, the question is to assign a vector state $|\psi\rangle\in\mathcal H$ to a resonance state. As it can be observed in several cases \cite{BOHM, BG}, its non-decay probability should follow an exponential law:

\begin{equation}\label{2}
\mathcal P(t) = |\langle\psi|e^{-itH}|\psi\rangle|^2 \varpropto e^{-\Gamma t/2}\,,\qquad t\ge 0\,.
\end{equation}

But not all models predict this behavior. Although for decaying states, $|\psi\rangle\in\mathcal H$, $\mathcal P(t)$ is approximately exponential for all times, there are large deviations from this exponential law for very short and very large values of time \cite{FGR}. These deviations are due to the semi-boundedness of the Hamiltonian. In fact, pure exponential decays exist if and only if the state $|\psi\rangle$ obeys a Breit Wigner energy distribution \cite{BOH}, something that  is not possible when the spectrum of $H$ is semi-bounded \cite{FGR,MS,KHA}. However, recent experiments suggest that, in some cases, there are indeed deviations from the exponential law \cite{Rothe, Urbanowski1, Bleistein}. For example, for short times, the probability stands approximatelly constant and decays slowly. For extremely large times a decay following a $t^{-1}$ law is observed. This regime, known as Khalflin effect, is very difficult to observe, although it has been experimentally detected \cite{Rothe,FGMR}. Nevertheless, for most of day life experiments the observational fact is that the non-decay probability is exponential, up to a reasonable degree of accuracy and for times of observation that are not too small to detect an effect or too large so that it just remains a quite small sample of undecayed products. Thus, for most experiments and  observational times the decay rule is a good approximation. If $|\psi\rangle\in\mathcal H$ is a decaying state, it admits the following decomposition:

\begin{equation}\label{3}
|\psi\rangle= |\psi^D\rangle + |\psi^B\rangle\,,
\end{equation}
where $|\psi^D\rangle$ decays exponentially for $t\ge 0$ and $|\psi^B\rangle$ represents the interaction with the environment, which is assumed to be responsible for all deviations of the pure exponential law. As $|\psi^D\rangle$ represents the ``intrinsic'' decay, it is often taken as the vector state for the resonance state. It is called the {\it decaying Gamow vector}. The requirement for the decaying Gamow vector, $|\psi^D\rangle$, to decay exponentially has motivated its definition proposed by Nakanishi \cite{NAK} as the eigenvector of the Hamiltonian with eigenvalue $z_R=E_R-i\Gamma/2$:

\begin{equation}\label{4}
H|\psi^D\rangle =z_R\,|\psi^D\rangle\,.
\end{equation}

As a consequence of this definition, one formally derives the exponential decay with time $t\ge 0$:

\begin{equation}\label{5}
e^{-itH}\,|\psi^D\rangle = e^{-itE_R}\,e^{-t\Gamma/2}\,|\psi^D\rangle\,, \qquad t\ge 0\,.
\end{equation}

This definition for the decaying Gamow vector has an apparent difficulty: If the Hamiltonian $H$ is self-adjoint in $\mathcal H$, it only admits real eigenvalues {\it with eigenvectors in $\mathcal H$}. However, it is possible to obtain complex eigenvalues if we define the eigenvalue equation in a larger space containing the Hilbert space as a subspace. This can be performed in the context of {\it rigged Hilbert spaces} (RHS). A RHS is a triplet of spaces \cite{GV,RO,ANT,MEL,GG,GG1}

\begin{equation}\label{6}
\Phi\subset \mathcal H \subset \Phi^\times\,,
\end{equation}
where $\mathcal H$ is an infinite dimensional separable Hilbert space. The dense subspace $\Phi$ has a topology finer than the topology inhereted from $\mathcal H$, so that the canonical injection $i:\Phi\longmapsto \mathcal H$, $i(\psi)=\psi$, $\forall\,\psi\in\Phi$, is continuous. The space $\Phi^\times$ is the antidual, i.e., the space of continuous antilinear functionals (mappings from $\Phi$ into the field of complex numbers $\mathbb C$) on $\Phi$. The action of $F\in\Phi^\times$ on $\phi\in\Phi$ is here denoted as $\langle\phi|F\rangle$, so as to preserve that Dirac bra-ket notation. The antilinearity of $F$ means that for all $\alpha, \beta\in\mathbb C$ and for all $\phi,\psi\in\Phi$, we have $\langle\alpha\phi+\beta\psi|F\rangle = \alpha^*\langle \phi|F\rangle+\beta^*\langle\psi|F\rangle$, where the star means complex conjugation. We endow $\Phi^\times$ with the weak topology with respect to the dual pair $\{\Phi,\Phi^\times\}$ \cite{RSI,HOR}. Each $\psi\in\mathcal H$ produces a unique $F_\psi\in\Phi^\times$ by $\langle \phi|F_\psi\rangle:=\langle \phi|\psi\rangle$, where $\langle \phi|\psi\rangle$ is the scalar product on $\mathcal H$.

Rigged Hilbert spaces have been used in order to implement the Dirac formalism of Quantum Mechanics \cite{BOHM,RO,ANT,MEL,GG,GG1}, for which the Hilbert space was not sufficient. Later, it was realized that Gamow vectors acquire full mathematical meaning in the context of RHS with an appropriate implementation \cite{BG,BOHM1, CG}. Also, RHS serve as a unifying framework for Lie group representation, special functions, discrete and continuous basis, with methods that include signal theory \cite{CGO,CGO1,CGO2}.

Since expressions like \eqref{4} and \eqref{5} are defined in the extension $\Phi^\times$ of the Hilbert space, we need a mechanism to extend the Hamiltonian and the evolution operator to $\Phi^\times$. This mechanism is the {\it duality formula}. It goes as follows: Let $\Phi\subset\mathcal H\subset \Phi^\times$ be a RHS and $A$  an operator (for us an operator is always linear and densely defined) on $\mathcal H$ verifying the following properties: i.) The adjoint of $A$, $A^\dagger$, preserves $\Phi$, i.e., for any $\varphi\in\Phi$, $A^\dagger\varphi\in\Phi$ (alternatively, we say that $A^\dagger\,\Phi\subset\Phi$); ii.) The adjoint $A^\dagger$ is continuous on $\Phi$ with respect to the topology that we have previously defined on $\Phi$. In this case, $A$ admits a unique continuous extension to $\Phi^\times$ given by the following duality formula:

\begin{equation}\label{7}
\langle A^\dagger\,\varphi|F\rangle = \langle\varphi|AF\rangle\,,\qquad \forall\,\varphi\in\Phi\,,\quad \forall\,F\in\Phi^\times\,.
\end{equation}

Note that $AF$ defines a unique element of $\Phi^\times$. If $A$ were Hermitian, \eqref{7} would have an obvious form. Furthermore, If $A$ is self-adjoint, there always exists a RHS fulfilling the above conditions with respect to $A$ \cite{RO}. In the case under discussion, being given the total Hamiltonian $H=H_0+V$, we may always construct a RHS, $\Phi_+\subset\mathcal H\subset\Phi_+^\times$, such that i.) $H\Phi_+\subset\Phi_+$; ii.) $H$ is continuous on $\Phi_+$; $H$ may be extended to $\Phi_+^\times$ with continuity; iv.) the eigenvalue expression $H|\psi^D\rangle= z_R\,|\psi^D\rangle$ makes full sense in $\Phi_+^\times$.

In addition, $e^{itH}\,\Phi_+\subset\Phi_+$ with continuity, at least for $t\ge 0$, so that its adjoint $e^{-itH}$ may be extended to $\Phi_+^\times$ by continuity, at least for $t\ge 0$. This implies that \eqref{5} makes full sense on $\Phi_+^\times$. As proven in \cite{BG,CG}, the space $\Phi_+$ admits a representation in terms of Hardy functions on the lower half of the complex plane.

We have already mentioned that resonance poles appear in complex conjugate pairs of the scattering matrix written in the energy representation. In addition to the pole located at $z_R=E_R-i\Gamma/2$, there exists the pole at $z_R^*=E_R+i\Gamma/2$. Then, there is another RHS $\Phi_-\subset\mathcal H\subset\Phi_-^\times$ and a vector $|\psi^G\rangle\in\Phi_-^\times$ with $|\psi^G\rangle\notin\mathcal H$, such that

\begin{equation}
H|\psi^G\rangle=z_R^*\,|\psi^G\rangle \,, \label{8} 
\end{equation}
\begin{equation}
e^{-itH}\,|\psi^G\rangle = e^{-itE_R}\,e^{t\Gamma/2}\,|\psi^G\rangle\,,\qquad t\le 0\,. \label{9}
\end{equation}

The space $\Phi_-$ admits a representation with Hardy functions on the upper half of the complex plane. The vector $|\psi^G\rangle$ is often called the {\it growing Gamow vector}. The decaying and the growing Gamow vectors are related via the time reversal operator $T$, which may be well defined on the RHS \cite{GM}, and verifies the following relations:

\begin{equation}\label{10}
T\,\Phi_\pm=\Phi_\mp\,,\quad T\,\Phi^\times_\pm=\Phi_\mp^\times\,, \quad T|\psi^D\rangle=|\psi^G\rangle\,, \quad T|\psi^G\rangle=|\psi^D\rangle\,.
\end{equation}

Then, for each resonance there are two state vectors, the growing and the decaying Gamow vectors. Both are equally suitable for a state vector and are just time reversal  of each other.

Finally, we wish to remark that we shall in principle assume that the resonance poles are simple. They may be multiple, and in that case $z_R$ and $z_R^*$ must have the same multiplicity \cite{AGP}. Models with double pole resonances have been constructed \cite{AGP,MH}.

\subsection{From non-commutativity to commutativity}

Along this subsection, we summarize a first approach to the transition from a commutative algebra of observables to a non-commutative one, taking the limit as $t\longmapsto\infty$. This cannot be achieved using a unitary evolution, so that we need to use a formalism in which time evolution is not unitary. Moreover, we have to make use of an structure different from the Hilbert space.

Leaving outside simple exactly solvable models such as the Friedrichs model \cite{F,EX,GP}, most quantum unstable models have a countable infinite number of resonances. Then, save for a finite number of resonances, either the resonance energy $E_R$ or the width $\Gamma$, or both, are too big. For big values of $E_R$ we leave the non-relativistic regime, so that we may discard these resonance poles. For large values of $\Gamma$, the mean life is so small that the resonance becomes unobservable. Therefore, in a first approximation, we may assume that the number of resonance poles is finite, no matter how large.

Thus, we assume that our system has $N$ resonances, with the resonance poles given by $z_1,z_2,\dots,z_N$ and their respective complex conjugates. The corresponding decaying Gamow vectors are

\begin{equation}\label{11}
|\psi_1^D\rangle, |\psi_2^D\rangle,\dots,|\psi_N^D\rangle\,,\qquad H|\psi_i^D\rangle=z_i\,|\psi_i^D\rangle \,,\quad i=1,2,\dots,N\,,
\end{equation}
and the corresponding growing Gamow vectors are

\begin{equation}\label{12}
|\psi_1^G\rangle, |\psi_2^G\rangle,\dots,|\psi_N^G\rangle\,,\qquad H|\psi_i^G\rangle=z_i^*\,|\psi_i^G\rangle \,,\quad i=1,2,\dots,N\,.
\end{equation}
Moreover, there are two generalized spectral decompositions of the Hamiltonian \cite{LFGH,GL}:

\begin{equation}\label{13}
H=\sum_{i=1}^N z_i\,|\psi_i^D\rangle\langle \psi_i^G| + {\rm BGR}\,,
\end{equation}
and

\begin{equation}\label{14}
H^\dagger=\sum_{i=1}^N z_i^*\,|\psi_i^G\rangle\langle \psi_i^D| + {\rm BGR}^*\,.
\end{equation}

The term BGR includes whatever does not depend on the Gamow vectors. This term is always non-vanishing.   Observe that decompositions \eqref{13} and \eqref{14} are the formal adjoint of each other. If $\Phi$ and $\Psi$ are two topological vector spaces, we denote by $\mathcal L(\Phi,\Psi)$ the space of all continuous linear operators from $\Phi$ into $\Psi$. Using this terminology and notation, one finds that \cite{GL}

\begin{equation}\label{15}
H\in \mathcal L(\Phi_-,\Phi_+^\times)\,,\qquad H^\dagger \in\mathcal L(\Phi_+,\Phi_-^\times)\,.
\end{equation}

Observe that $H$ is an operator on $\Phi_-$. The action of $H$ on a vector of $\Phi_-$ gives a vector in $\Phi_+^{\times}$. A similar comment is in order about $H^\dagger$ as a linear mapping from $\Phi_+$ into $\Phi_-^\times$. Since the description of resonances may be given in terms of decaying and growing Gamow vectors, it results that the distinction between $H$ and $H^\dagger$ is purely conventional and these two operators represent two similar {\it non-hermitic} decomposition of the Hamiltonian which are formally time reversal of each other.  However, the action of \eqref{13} or \eqref{14} on the Gamow vectors has no sense as Gamow vectors do not belong neither to $ \Phi_-$ nor to $\Phi_+$. We need to give a meaning to such actions.

We are interested in what concerns to resonances and, in consequence, we drop the background terms.  Then, both $H$ and $H^\dagger$ are non-hermitian operators depending only on resonance poles and Gamow vectors, although still verify \eqref{15}. These {\it truncated} versions of $H$ and $H^\dagger$ are as in \eqref{13} and \eqref{14}, but without the BGR term.

Let us consider the $2N$ dimensional space $\mathcal H^G$, spanned by the following vectors:

\begin{equation}\label{16}
\{|\psi^D_1\rangle\,,|\psi^G_1\rangle\,,|\psi^D_2\rangle\,,|\psi^G_2\rangle\,,\dots\,,|\psi^D_N\rangle\,,|\psi^G_N\rangle\}\,.
\end{equation}

Then, we define a pseudometric on $\mathcal H^G$ as the bilinear form given in the basis \eqref{16} by the following matrix:

\begin{equation}\label{17}
A:= \left(\begin{array}{ccccccc} 0 & 1 & \dots & \dots & \dots & \dots & \dots \\ 1 & 0 & \dots & \dots & \dots & \dots & \dots \\ \dots & \dots & 0 & 1 & \dots & \dots & \dots  \\  \dots & \dots & 1 & 0 & \dots & \dots & \dots \\ \dots & \dots & \dots & \dots & \dots & \dots & \dots   \\ \dots & \dots & \dots & \dots & \dots & 0 & 1 \\ \dots & \dots & \dots & \dots & \dots & 1 & 0  \end{array}\right)\,.
\end{equation}
All entries replaced by dots are equal to zero.
Then, the psuedo-scalar product of two vectors $|\psi\rangle,|\varphi\rangle\in\mathcal H^G$ is defined as

\begin{equation}\label{18}
(\psi|\varphi) := \langle \psi|A|\varphi\rangle\,,
\end{equation}
so that

\begin{equation}\label{19}
(\psi^D_i|\psi^D_j)=(\psi^G_i|\psi^G_j)=0\,, \quad (\psi^D_i|\psi^G_j)=(\psi_i^G|\psi_j^D)=\delta_{ij}\,.
\end{equation}

This pseudometric is important in this context in order to define the operations that are relevant for the construction of an evolution operator.  Assume that we define a standard scalar product on $\mathcal H^G$. The simplest one implies that \eqref{16} is an orthonormal basis. For simplicity, we may assume first the presence of only one resonance. Then,

\begin{equation}\label{20}
H^2=z_R^2 |\psi^D\rangle\langle\psi^G|\psi^D\rangle\langle\psi^G|=0.
\end{equation}

Therefore, to avoid this kind of problems maintaining some simplicity in the operations is why we introduce this pseudometric. From \eqref{18}, we note that if $B$ is a matrix such that $A=B^2$, one has that $B|\psi_i^D\rangle\equiv |\psi_i^D)$ and $\langle\psi_i^G|B\equiv (\psi_i^G|$, $i=1,2,\dots,N$.  The matrix $B$ is not uniquely defined, although a reasonable choice is given by replacing the $2\times 2$ dimensional non-vanishing boxes in \eqref{17} by

\begin{equation}\label{21}
B=(-i)^{1/2} \left(\begin{array}{cc} i\sqrt 2/2 & \sqrt 2/2 \\[2ex] \sqrt 2/2 & i\sqrt 2/2  \end{array}\right)\,.
\end{equation}

With some natural amendments in the formalism, $|\psi^D)$ and $|\psi^G)$ will represent the Gamow vectors. Then, we have to replace \eqref{13}, without the BGR term, by

\begin{equation}\label{22}
H=\sum_{i=1}^N z_i\,B|\psi_i^D\rangle\langle \psi_i^G|B= \sum_{i=1}^N z_i\,|\psi_i^D)(\psi_i^G|  \,,
\end{equation}
so that

\begin{equation}\label{23}
H|\psi_j^D) = \sum_{i=1}^N z_i\,|\psi_i^D)(\psi_i^G|\psi_j^D)= \sum_{i=1}^N z_i\,|\psi_i^D)\,\delta_{ij} = z_j\, |\psi_j^D)\,,
\end{equation}

\begin{equation}\label{24}
H^2 = B\left[ \sum_{i=1}^N z_i^2\,|\psi_i^D\rangle\langle \psi_i^G| \right]B = \sum_{i=1}^N z_i^2\,|\psi_i^D)(\psi_i^G|\,.
\end{equation}

We obtain in the same way $H^n$, so that we arraive to the following formal expression:

\begin{equation}\label{25}
e^{-itH} = \sum_{j=1}^N e^{-itz_j} \,|\psi_j^D)(\psi^G_j|\,.
\end{equation}

If we replace $B$ by $C:=B^\dagger$, we have $|\psi_i^G)\equiv C|\psi_i^G\rangle$ and $(\psi_i^D|C\equiv \langle\psi^D|$. Then, $H^\dagger$ becomes:

\begin{equation}\label{26}
H^\dagger= C\left[\sum_{i=1}^N z_i^* \,|\psi_i^G\rangle\langle\psi_i^D| \right]C = \sum_{i=1}^N z_i^* \,|\psi_i^G)(\psi_i^D|\,,
\end{equation}
so that

\begin{equation}\label{27}
H^\dagger\,|\psi_i^G)=z_i^*\,|\psi_i^G)\,,\qquad e^{-itH^\dagger} = \sum_{j=1}^N e^{-itz_j^*} \,|\psi_j^G)(\psi_j^D|\,.
\end{equation}

In addition, we have the following results:

\begin{eqnarray}\label{28}
&&H|\psi_j^G) =0\,,\qquad H^\dagger|\psi_j^D)=0\,,\quad j=1,2,\dots,N \nonumber\\[2ex] &&e^{-itH}|\psi_j^D) = e^{-itz_j}\,|\psi_j^D)\,,\qquad e^{-itH^\dagger}|\psi_j^G) = e^{-itz^*_j}\,|\psi_j^G)\,,\nonumber\\[2ex]  &&e^{-itH}|\psi_j^G)=0\,,\qquad e^{-itH^\dagger}|\psi_j^D)=0\,.
\end{eqnarray}

Therefore, both evolution operators $e^{-itH}$ and $e^{-itH^\dagger}$ act non-trivially on a half of the space $\mathcal H^G$ and vanish on the other half. However, it is desirable that meaningful operators on resonance states act on all Gamow states. Therefore, we need an obvious extension of the above formalism to all $\mathcal H^G$. To such end, we define the Hamiltonian as

\begin{equation}\label{29}
H= \sum_{i=1}^N z_i\,|\psi_i^D)(\psi_i^G|+z_i^*\,|\psi_i^G)(\psi_i^D|\,.
\end{equation}
With this definition, $H$ is formally Hermitian. Using the pseudometric, we obtain the following properties:

\begin{equation}\label{30}
H|\psi_i^D)=z_i\,|\psi_i^D)\,,\qquad H|\psi_i^G)=z_i^*\,|\psi_i^G)\,,
\end{equation}

\begin{equation}\label{31}
H^n= \sum_{i=1}^N z^n_i\,|\psi_i^D)(\psi_i^G|+(z_i^*)^n\,|\psi_i^G)(\psi_i^D|\,,
\end{equation}
so that formally:

\begin{equation}\label{32}
U(t):= e^{-itH} = \sum_{j=1}^N e^{-itz_j}\,|\psi_j^D)(\psi_j^G| + e^{-itz_j^*}\,|\psi_j^G)(\psi^D|\,.
\end{equation}
The identity operator admits the following decomposition:

\begin{equation}\label{33}
I=\sum_{i=1}^N \{\,|\psi^D_i)(\psi_i^G|+|\psi_i^G)(\psi_i^D|\,\}\,.
\end{equation}
Note that from \eqref{32}, we have that

\begin{equation}\label{34}
U(-t)= e^{itH} = \sum_{j=1}^N e^{itz_j}\,|\psi_j^D)(\psi_j^G| + e^{itz_j^*}\,|\psi_j^G)(\psi^D|\,.
\end{equation}

Hence, using once again the pseudometric relations \eqref{19}, we find that $U(t)U(-t)=U(-t)U(t)$ $=I$, so that $U(-t)=U^{-1}(t)$. So far, everything sounds like ordinary quantum mechanics. If we proceed further, the time evolution of an arbitrary linear operator $O=O(0)$ should be given by $O(t)=U(-t)\,O\,U(t)$. However, this produces a contribution in $O(t)$ that grows exponentially, which cannot be  cancelled out with any other term \cite{LFGH}. This suggests that \eqref{32} may not be a good definition for the time evolution operator in our case. For this reason, we propose the following definition for the time evolution operator:

\begin{equation}\label{35}
U(t) = e^{-itz_R}\,|\psi^D)(\psi^G| + e^{itz_R^*}\,|\psi^G)(\psi^D|\,,
\end{equation}
which is formally Hermitic. This means that the formal adjoint is given by $U^\dagger(t)=U(t)$. In this case, the relation $U^\dagger(t)U(t)=I$ is not longer correct. Instead we have

\begin{equation}\label{36}
U(t)U^\dagger(t)=U^2(t)=e^{-t\Gamma}\,I\,.
\end{equation}

This is not the usual quantum relation, although the exponential decay is consistent with the exponential decay of Gamow states. Observe that \eqref{35} is already time asymmetric. For the time evolution of observables, one should choose $O(t)=U^\dagger(t)\,O\,U(t)$, which is consistent with the usual Heisenberg picture.

Then, let us take two arbitrary operators $O_1$ and $O_2$ on $\mathcal H^G$, let them evolve with time and construct their commutator at time $t$, which gives \cite{LFGH}

\begin{equation}\label{37}
[O_1(t),O_2(t)] = \sum_{i=1}^N e^{-t\Gamma_i}\, \{\alpha_i(t)\,|\psi_i^D)(\psi_i^G|+\beta_i(t)\,|\psi_i^G)(\psi_i^D|\}\,,
\end{equation}
where $\alpha_i(t)$ and $\beta_i(t)$, $i=1,2,\dots,N$, are oscillating functions of time.

\subsection{Time reversal operation}

It is interesting to discuss the previous formalism under the time reversal operation. We have previously established that if $T$ is the time reversal operator, $T|\psi^D\rangle=|\psi^G\rangle$ and $T|\psi^G\rangle=|\psi^D\rangle$. Then, one may expect that the same operations result when we use $|\psi^D)$ and $|\psi^G)$ instead. However, this is not true, unless we use a correct definition for the time reversal operator suitable for this case. In fact, Wigner showed that there are four distinct definitions of the time reversal operator \cite{GM,W,WI}. Although these definitions were, in principle, related with some projective representations of the Poincar\'e group extended with time inversion and parity, we shall see that one of these two dimensional choices for $T$ is suitable for our discussion.

For simplicity, we assume that the number of resonances is $N=1$. The extension to an arbitrary number of resonances is straightforward. Now, we adopt the following notation for the basis of $\mathcal H^G$ as $|\psi^D\rangle=\left(\begin{array}{c} 1\\0\end{array} \right)$ and $|\psi^G\rangle=\left(\begin{array}{c} 0\\1\end{array} \right)$. Let us choose as time reversal operator \cite{GM,W,WI} the following matrix, with $C$ the complex conjugation operation:

\begin{equation}\label{38}
T:= \left(\begin{array}{cc} 0 & \mathcal C \\[2ex] \mathcal C & 0\end{array}\right)\,.
\end{equation}
Then,

\begin{eqnarray}\label{39}
T|\psi^D)= TB \,\left(\begin{array}{c} 1\\0\end{array} \right) = \left(\begin{array}{cc} 0 & \mathcal C \\[2ex] \mathcal C & 0\end{array}\right) \, (-i)^{1/2}\left(\begin{array}{cc} i\sqrt 2/2 & \sqrt 2/2 \\[2ex] \sqrt 2/2 & i\sqrt 2/2 \end{array}\right) \left(\begin{array}{c} 1\\[2ex] 0\end{array} \right)=\nonumber  \\[2ex] = i^{1/2} \left(\begin{array}{c} \sqrt 2/2 \\[2ex] -i\sqrt 2/2 \end{array}\right) = i^{1/2} \left(\begin{array}{cc} -i\sqrt 2/2 & \sqrt 2/2 \\[2ex] \sqrt 2/2 & -i\sqrt 2/2\end{array}\right)\left(\begin{array}{c} 0\\[2ex] 1\end{array} \right) = B^\dagger |\psi^G\rangle =|\psi^G)\,.
\end{eqnarray}
Analogously,

\begin{equation}\label{40}
T|\psi^G)=|\psi^D)\,.
\end{equation}

However, the Hamiltonian \eqref{29} is not time reversal invariant. It may be shown by a simple calculation. Let $|\psi)=\sum_{i=1}^N a_i\,B|\psi^D_i\rangle + \sum_{i=1}^N b_i\,B^\dagger\,|\psi^G\rangle$ be an arbitrary vector in $\mathcal H^G$, where $a_i$ and $b_i$ ($i=1,\dots,N$) are complex numbers. We need to compare

\begin{equation}\label{41}
(\psi|H|\psi)= \sum_{i=1}^N z_i (\psi|\psi^D_i)(\psi_i^G|\psi) + \sum_{i=1}^N z_i^* (\psi|\psi^G_i)(\psi^D_i|\psi)\,,
\end{equation}
with

\begin{equation}\label{42}
(\psi|THT|\psi)= \sum_{i=1}^N z^*_i (\psi|T|\psi^D_i)(\psi_i^G|T|\psi) + \sum_{i=1}^N z_i (\psi|T|\psi^G_i)(\psi^D_i|T|\psi)\,.
\end{equation}

This calculation is rather straightforward and we sketch it in Appendix. The conclusion is that \eqref{41} and \eqref{42} do not coincide and therefore, the model is not time reversal invariant.

\section{Algebraic formalism}

In this section, we will use an alternative formalism for the description of observables and states of the non-relativistic quantum mechanics, which was originally proposed in order to include the van Hove states that apear in systems far from thermodynamic equilibrium \cite{ALST}. This formalism was extended to further purposes \cite{CAS, CAS2}. In particular, it has been useful to accommodate Gamow states for resonances as functionals over some type of algebras. The presentation of quantum states as functionals on algebras generated by observables is a common feature. Thus, we have shown that Gamow states for quantum unstable systems are as valid as any other type of states.

Our aim is to use this formalism in order to describe the transition of the algebra of observables from non-commutativity to commutativity. This is not particularly new, although no emphasis in this fact has been given so far. For our discussion, we will follow the formalism described in \cite{CGIL}. But there the purpose was different, the aim was to include Gamow states as functionals on certain algebras of observables. Note that in \cite{CGIL}, we have focused our attention in the behavior of states. Now, we will focus on the time evolution of the observables, so that we shall use a type of Heisenberg time evolution.
We have discussed the motivation for the formalism given in the present section in \cite{CGIL}; therefore, we shall not insist on the motivation and the details in this work.

Resonances are typically produced by a Hamiltonian pair $\{H_0,H=H_0+V\}$, and we assume that it has the simplest properties and works in the energy representation. In particular, the spectrum of the {\it free} Hamiltonian $H_0$ is purely absolutely continuous, simple and given by $\mathbb R^+\equiv [0,\infty)$. Let $|E\rangle$ be the eigenvector of $H_0$ for the particular value of the energy $E\in\mathbb R^+$. The meaning of such eigenvectors has been largely discussed as functionals on certain rigged Hilbert space. They are also valid for the expansion of $H_0$ in terms of generalized projections \cite{AGS, GA} as

\begin{equation}\label{3.1}
H_0= \int_0^\infty dE\,E \,|E\rangle\langle E|\,.
\end{equation}

An operator $O$ is said to be compatible with $H_0$ if it has the form

\begin{equation}\label{3.2}
O= \int_0^\infty dE\, O(E)\,|E\rangle\langle E| + \int_0^\infty dE \int_0^\infty dE\, O(E,E')\,|E\rangle\langle E'|\,,
\end{equation}
where $O(E)$ and $O(E,E')$ are given well behaved functions, which always can be multiplied by each other. From this point of view, they could be considered as test functions. In general $O(E,E')\ne O(E',E)$. One of the properties of the kets $|E\rangle$ is that

\begin{equation}\label{3.3}
\langle E|E'\rangle =\delta(E-E')\,.
\end{equation}

A proper choice of the space of test functions shows that the set of operators which are compatible with $H_0$ is a non-commutative algebra with identity, $\mathcal A_0$. The proof is straightforward. The identity is given by

\begin{equation}\label{3.4}
I= \int_0^\infty dE\,|E\rangle\langle E|\,.
\end{equation}

Next, we consider the scattering produced by the Hamiltonian pair $\{H_0,H\}$. Some extra assumptions are needed, which we choose with criteria of simplicity and generality. We  assume asymptotic completeness and the existence of the M{\o}ller wave operators

\begin{equation}\label{3.5}
\Omega_\pm:= \lim_{t\to \pm} e^{itH}\,e^{-itH_0} \,\varphi=: \varphi^\pm\,,
\end{equation}
with $\varphi$ a state that evolves freely. Thus, the M{\o}ller wave operators relate states that evolve freely, according to the hamiltonian $H_0$, with states $\varphi^\pm$, that evolve under the total Hamiltonian $H=H_0+V$, where $V$ is a potential. Note that $\lim_{t\to\pm}(e^{-itH_0}\,\varphi-e^{-itH}\,\varphi^\pm)=0$\,.

We define the following states:

\begin{equation}\label{3.6}
|E^\pm\rangle =\Omega_\pm\,|E\rangle\,,
\end{equation}
which are eigenvectors of $H$, with eigenvalue $E\in\mathbb R^+$, i.e.,  $H|E^\pm\rangle = E\,|E^\pm\rangle$. States given in equation \eqref{3.6} have a proper meaning \cite{CGIL}. Since \eqref{3.6} implies that $\langle E|\Omega_\pm^\dagger = \langle E^\pm|$, from the following  definition

\begin{equation}\label{3.7}
O^\pm := \Omega_\pm \,O\, \Omega_\pm^\dagger\,,
\end{equation}
we have that

\begin{equation}\label{3.8}
O^\pm = \int_0^\infty dE\, O(E)\,|E^\pm\rangle\langle E^\pm|+ \int_0^\infty dE \int_0^\infty dE'\,O(E,E')\,|E^\pm\rangle\langle {E'}^\pm|\,.
\end{equation}
We say that an operator $O$ is {\it compatible with} $H$ if it has the form \eqref{3.8} with either sign.

Taking into account that

\begin{equation}\label{3.9}
\langle E^\pm|w^\pm\rangle = \langle E|\,\Omega^\dagger_\pm \,\Omega_\pm\,|w\rangle =\langle E|w\rangle =\delta(E-w)\,,
\end{equation}
we conclude that the operators of the type $O^+$ and $O^-$ in \eqref{3.6} form  algebras $\mathcal A_+$ and $\mathcal A_-$, respectively. These algebras have the following identities,

\begin{equation}\label{3.10}
I_\pm = \int_0^\infty dE\,|E^\pm\rangle\langle E^\pm|\,.
\end{equation}

The relation between the unperturbed algebra $\mathcal A_0$ and $\mathcal A_\pm$ is the following

\begin{equation}\label{3.11}
\mathcal A_\pm =\Omega_\pm\,\mathcal A_0\,\Omega^\dagger_\pm\,.
\end{equation}

we also introduce the notation  $|E^\pm)$ for $|E^\pm\rangle\langle E^\pm|$ and $|E^\pm {E'}^\pm)$ for $|E^\pm\rangle\langle {E'}^\pm|$, so that \eqref{3.8} takes the form given by

\begin{equation}\label{3.12}
O^\pm = \int_0^\infty dE\,O(E)\,|E^\pm) + \int_0^\infty dE \int_0^\infty dE'\,O(E,E')\,|E^\pm {E'}^\pm)\,.
\end{equation}
Observe that, using this notation,  $(E^\pm|O^\pm)=O(E)$ and that $(E^\pm {E'}^\pm|O^\pm)=O(E,E')$. As a simple remark, $O^\pm$ is an observable if and only if $O(E)=O^*(E)$ and $O(E,E')=O^*(E',E)$, where the star denotes complex conjugation \cite{CGIL}.

Now, we are going to consider the time evolution in the Heisenberg picture. Recall that $H|E^\pm\rangle =E\,|E^\pm\rangle$, which implies that

\begin{equation}\label{3.13}
e^{itH_0}\,|E^\pm\rangle \langle {E'}^\pm|\,e^{-itH_0} = e^{it(E-E')}\,|E^\pm\rangle \langle {E'}^\pm|\,.
\end{equation}

Expression \eqref{3.13} is useful in order to determine the time evolution of the elements of the algebras $\mathcal A_\pm$ in the Heisenberg picture 

\begin{eqnarray}\label{3.14}
O^\pm(t):= e^{itH_0}\,O^\pm\,e^{-itH_0} \nonumber\\[2ex] = \int_0^\infty dE\,O(E) \,|E^\pm\rangle\langle E^\pm | + \int_0^\infty dE \int_0^\infty dE' \,O(E,E')\, e^{it(E-E')}\,|E^\pm\rangle\langle {E'}^\pm| \nonumber\\[2ex]
= \int_0^\infty dE\,O(E) \,|E^\pm) + \int_0^\infty dE \int_0^\infty dE' \,O(E,E')\, e^{it(E-E')}\, |E^\pm{E'}^\pm)  \,.
\end{eqnarray}

We are going to analyze the limits $t\longmapsto\pm\infty$ of $O^\pm(t)$, respectively. This requires the concept of functional on the algebras $\mathcal A_\pm$. Each functional on $\mathcal A_\pm$ can be formally written as

\begin{equation}\label{3.15}
\rho_\pm =\int_0^\infty dE\,\rho(E) \,(E^\pm|+ \int_0^\infty dE \int_0^\infty dE'\,\rho(E,E') \,(E^\pm{E'}^\pm|\,,
\end{equation}
where $\rho(E)$ and $\rho(E,E')$ are functions or generalized functions acting on the space of test functions $O(E)$ and $O(E,E')$, respectively. Also, the functionals $(E^\pm|$ and $(E^\pm{E'}^\pm|$ are defined as

\begin{equation}\label{3.16}
(E^\pm |O^\pm)=O(E)\,,\qquad (E^\pm{E'}^\pm|O^\pm)=O(E,E')\,,
\end{equation}
and have the following properties:

\begin{eqnarray}\label{3.17}
(E^\pm|w^\pm)=\delta(E-w)\,,\qquad (E^\pm|w^\pm{E'}^\pm)=0\,, \nonumber\\[2ex] \qquad (E^\pm{E'}^\pm|w^\pm{w'}^\pm) = \delta(E-w)\delta(E'-w')\,.
\end{eqnarray}
The action of $\rho_\pm$ on $O^\pm\in\mathcal A_\pm$ is given by

\begin{equation}\label{3.18}
(\rho^\pm|O^\pm)= \int_0^\infty dE\,\rho(E)\,O(E) + \int_0^\infty dE \int_0^\infty dE' \,\rho(E,E')\,O(E,E')\,,
\end{equation}
so that

\begin{eqnarray}\label{3.19}
(\rho^\pm|O^\pm(t)) = \int_0^\infty dE\,\rho(E)\,O(E) + \int_0^\infty dE \int_0^\infty dE' \, e^{it(E-E')}\, \rho(E,E')\,O(E,E')\,.
\end{eqnarray}

Often, the product $\rho(E,E')\,O(E,E')$ is an integrable function. In this case, the Riemann-Lebesgue theorem shows that the last term of \eqref{3.19} vanishes as $t\longmapsto\pm\infty$. Therefore, we may say that

\begin{equation}\label{3.20}
\lim_{t\to\pm\infty} O^\pm(t) =\int_0^\infty dE\,O(E)\,|E^\pm)\,,
\end{equation}
in a weak sense. This is a mathematical fact that has been named as {\it self-induced decoherence} in the literature \cite{CAS, CAS2}. What is interesting from our point of view is that if $O^\pm$ and $U^\pm$ are two operators in the algebra $\mathcal A_\pm$, one can obtain the following result

\begin{equation}\label{3.21}
\lim_{t\to\pm\infty} [O^\pm(t),U^\pm(t)] = 0\,.
\end{equation}
Note that in general $[O^\pm(t),U^\pm(t)] \ne 0$ for all finite times. 
A similar mathematical result has also been studied by Kiefer and Polarski \cite{Kiefer}, and Ram\'irez and Reboiro \cite{Reboiro, Reboiro2}, using alternative approaches.

Therefore, we have obtain the same result found in the previous section, a quantum-to-classical transition for very large values of time. Nevertheless, there are some fundamental differences between the discussion on Section 2 and the presentation in this section. The most important one is that we do not make use of resonance states here. In addition, we do not mix in and out states in the same framework. Quite the contrary, the algebras $\mathcal A_-$ and $\mathcal A_+$, which refers to in and out observables are treated separately. In spite of this terminology of in and out observables, the algebras $\mathcal A_\pm$ are just time reversal of each other as demonstrated in \cite{CGIL}.

\section{Conclusions}


Non-commutativity of observables is one of the characteristic features of quantum mechanics and it is related to the impossibility of realizing simultaneous measurements of incompatible observables.
On the contrary, commutativity of observables is related with a classical behavior, in the sense that simultaneous measurements of compatible observables are always possible.
The transition from a non-commutative algebra of observables to commutative one is a fundamental aspect of the quantum-to-classical limit.
However, if we only consider unitary evolutions, it is not possible to describe this transition.
In order to describe the quantum-to-classicaltransition, we need to consider more general time evolutions.
One way to do that is to consider the Gamow vectors, which represent resonances including growing and decaying states, and they are linked with exponential decays in the evolution.

In this paper, we gave a further generalization of the non-unitary
evolution presented in previous papers. 
We discussed two quantum models
in which the 
transitions from the non-commutativity to commutativity happens when time goes to infinite. In the first one, the involved operators  act on a space spanned by the Gamow states, endowed with a Krein pseudo-metric. Taking into account that resonances decay as time tends to infinity, we have to use a proper definition of time evolution on vectors of the Gamow space compatible with this fact. Then, all observables in this model commute when time goes to infinite.

In the second one, we use an algebraic formalism from scattering theory. We
construct in and out algebras of non-commuting operators, that commute in the limits $t\longmapsto\pm\infty$ in a weak sense.
These algebras has been constructed in order to include certain states representing situations far from equilibrium. It has also been used for describing Gamow states as functionals over these two algebras and for a formulation of the decoherence.

\section*{Acnowledgements}

M. Gadella acknowledges partial financial support to the Spanish Government Grant MTM2014-57129-C2-1-P,  the Junta de Castilla y Le\'on Grants BU229P18, VA137G18. S. Fortin, F. Holik and M. Losada wish to acknowledge the financial support
of the Universidad de Buenos Aires, the grant PICT-2014-2812 from the Consejo Nacional de Investigaciones
Cient\'ificas y T\'ecnicas of Argentina.

\section*{Appendix}

Let us show that formulas \eqref{41} and \eqref{42} are not equivalent. For simplicity, we shall assume that there is only one resonance so that dim $\mathcal H^G=2$. The vector $|\psi)$ being arbitrary is a linear combination of $|\psi^D\rangle$ and $|\psi^G\rangle$, so that it may be written as a column vector as $\left(\begin{array}{c} a \\ b\end{array}\right)$ with $a$ and $b$ complex. Then,

\begin{eqnarray}
(\psi^D|\psi)= (1,0) B^\dagger\, \left(\begin{array}{c} a \\[2ex] b\end{array}\right)=  (1,0) \,i^{1/2} \left(\begin{array}{cc} -i\sqrt 2/2 & \sqrt 2/2 \\[2ex] \sqrt 2/2 & -i\sqrt 2/2\end{array}\right)\left(\begin{array}{c} a \\[2ex] b\end{array}\right) \nonumber\\[2ex] =i^{1/2} \left[\frac{\sqrt 2}{2}\,b -i\,\frac{\sqrt 2}{2}\,a\right]\,.
\end{eqnarray}
Similar calculations yield (we have written $\sqrt{-1}=-i$):

\begin{equation}
(\psi^G|\psi) = (\psi^D|\psi)\,, \quad (\psi|\psi^D)=(\psi|\psi^G)= i^{1/2} \left[\frac{\sqrt 2}{2}\,b^* -i\,\frac{\sqrt 2}{2}\,a^* \right]\,.
\end{equation}
This gives \eqref{41}. To obtain \eqref{42}, we need the following calculation:
\begin{eqnarray}
(\psi^D|T|\psi)= (1,0) \, i^{1/2} \left(\begin{array}{cc} -i\sqrt 2/2 & \sqrt 2/2 \\[2ex] \sqrt 2/2 & -i\sqrt 2/2\end{array}\right) \left(\begin{array}{cc} 0 & C \\[2ex] C & 0 \end{array}\right) \left(\begin{array}{c} a \\[2ex] b\end{array}\right) \nonumber\\[2ex] = i^{1/2}\, \left(-i\,\sqrt 2/2, \sqrt 2/2 \right) \left(\begin{array}{c} b^* \\[2ex] a^*\end{array}\right) = i^{1/2}\, \left[\frac{\sqrt 2}{2}\,a^* -i\, \frac{\sqrt 2}{2}\,b^* \right]\,.
\end{eqnarray}

Taken $\sqrt{-1}=-i$, we easily find that $(\psi^D|T|\psi)=(\psi^G|T|\psi)$. From the other terms, we write

\begin{eqnarray}
(\psi|T|\psi^D) = (a^*,b^*) \left(\begin{array}{cc} 0 & C \\[2ex] C & 0 \end{array}\right) (-i)^{1/2} \left(\begin{array}{cc} i\sqrt 2/2 & \sqrt 2/2 \\[2ex] \sqrt 2/2 & i\sqrt 2/2\end{array}\right) \left(\begin{array}{c} 1 \\[2ex] 0\end{array}\right) \nonumber\\[2ex] = i^{1/2} (a^*,b^*) \left(\begin{array}{cc} 0 & C \\[2ex] C & 0 \end{array}\right) \left(\begin{array}{c}\sqrt 2/2 \\[2ex] -i\sqrt 2/2 \end{array}\right) = i^{1/2}\, \left[\frac{\sqrt 2}{2}\,a^* -i\, \frac{\sqrt 2}{2}\,b^* \right]\,.
\end{eqnarray}
Similarly, we obtain that
\begin{equation}
\label{ec apen final}
(\psi|T|\psi^G)=(\psi^G|T|\psi)=(\psi^D|T|\psi)=(\psi|T|\psi^D)\,.
\end{equation}
The obvious conclusion is that \eqref{41} and \eqref{42} do not coincide. The first and third identities in \eqref{ec apen final} are not a surprise due to the properties of the time reversal operator $T$.

\section{Compliance with ethical standards}

\noindent Conflict of interest: The authors declare that they have no conflict of interest.

\noindent Ethical approval: This article does not contain any studies with human participants or animals performed by any of the authors.


\begin{thebibliography}{99}

\bibitem{FV} S. Fortin, L. Vanni, Found. Phys., {\bf 44} (2014) 1258-1268.

\bibitem{Fortin2013} M. Castagnino and S. Fortin, Formal features of a General Theoretical Framework for Decoherence in open and closed systems, Int. J. Theor. Phys., \textbf{52} (2013) 1379-1398.

\bibitem{Schlo}
M. Schlosshauer, \textit{Decoherence and the Quantum-To-Classical Transition}, Springer, Berlin, 2007.

\bibitem{Fortin2010} M. Castagnino, S. Fortin and O. Lombardi, The effect of random coupling coefficients on decoherence, Modern Physics Letters A, \textbf{25}  
(2010) 611-617.




\bibitem{LFH} M. Losada, S. Fortin, F. Holik, Int. J. Theor. Phys., {\bf 57} (2018) 465-475.

\bibitem{LFGH} M. Losada, S. Fortin, M. Gadella, F. Holik, Int. J. Mod. Phys. A, {\bf 33}, 18-19 (2018) 1850109.

\bibitem{FHV} S. Fortin, F. Holik, L. Vanni, 
Non-unitary evolutions in quantum logics, in F. Bagarello, R. Passante, C. Trapani (eds) Non-Hermitian Hamiltonians in Quantum Physics. Springer Proceedings in Physics, \textbf{184}. Springer, Cham 2016.


\bibitem{BEU} A. Bohm, F. Erman, H. Uncu, Turk. J. Phys., {\bf 35} (2011) 209-240.



\bibitem{FGR} L. Fonda, G.C. Ghirardi, A. Rimini, Rep. Progr. Phys., {\bf 41} (1978) 587-631.

\bibitem{BOH} A. Bohm, {\it Quantum Mechanics: Foundations and Applications}, Springer, Berlin, 1993.

\bibitem{GAD} M. Gadella, Quantum resonances: Theory and models, in  P. Kielanowski et (eds) Geometric Methods in Physics, XXXII Workshop Bialowieza.
Springer Bassel AG, Poland 2014, 99-118.
 

\bibitem{GKN} M. Gadella, {\c S}. Kuru, J. Negro, Ann. Phys., {\bf 379} (2017) 86-101.

\bibitem{RSIV} M. Reed, B. Simon, {\it Analysis of Operators}, Academic Press, New York, 1978.

\bibitem{NUS} H.M. Nussenzveig, {\it Causality and Dispersion Relations}, Academic Press, New York and London, 1972.

\bibitem{F} K.O. Friedrichs, Commun. Appl. Math., {\bf 1} (1948) 361-406.

\bibitem{EX} P. Exner, {\it Open Quantum Systems and Feynman Integrals}, Reidel, Dordrecht 1984.

\bibitem{GP} M. Gadella, G.P. Pronko, Fort. Phys., {\bf 59} (2011) 795-859.

\bibitem{BOHM} A. Bohm, \textit{The Rigged Hilbert Space and Quantum Mechanics}. Springer Lecture Notes in Physics, \textbf{78}, Springer, New York 1978.


\bibitem{BG} A. Bohm, M. Gadella, \textit{Dirac Kets, Gamow Vectors and Gelfand Triplets}. Springer Lecture Notes in Physics, \textbf{348}, Springer, New York and Berlin, 1989.

\bibitem{MS} B. Misra, E.C.G. Sudarshan, J. Math. Phys., {\bf 18} (1977) 756-763.

\bibitem{KHA} L.A.  Khalfin, JETP Lett., {\bf 15} (1972) 388-392.

\bibitem{Rothe} C. Rothe, S. I. Hintschich, and A. P. Monkman, Phys. Rev. Lett., {\bf 96}  (2006) 163601.

\bibitem{Urbanowski1}     K. Urbanowski, The European Physical Journal D, {\bf 54} (2009) 25-29 .


\bibitem{Bleistein} N. Bleistein, R. Handelsman, \textit{Asymptotic expansion of integrals}. Dover Inc., New York, 1986.


\bibitem{FGMR} M.C. Fischer, B. Guti\'errez-Medina, M.G. Reizen, Phys. Rev. Lett., {\bf 87} (2001) 40402.


\bibitem{NAK} N. Nakanishi, Progr. Theor. Phys., {\bf 19} (1958) 607-621.

\bibitem{GV} I.M. Gelfand, N.Y. Vilenkin, \textit{Generalized Functions: Applications to Harmonic Analysis}. Academic, New York, 1964.

\bibitem{RO} J.E. Roberts, Commun. Math. Phys., {\bf 3} (1966) 98-119.

\bibitem{ANT} J.P. Antoine, J. Math. Phys., {\bf 10} (1969) 53-69.

\bibitem{MEL} O. Melsheimer, J. Math. Phys., {\bf 15} (1974) 902-916.

\bibitem{GG} M. Gadella, F. G\'omez, Found. Phys., {\bf 32} (2002) 815-869.

\bibitem{GG1} M. Gadella, F. G\'omez,, Int. J. Theor. Phys., {\bf 42} (2003) 2225-2254.

\bibitem{RSI} M. Reed, B. Simon, \textit{Functional Analysis}, Academic, New York, 1981.

\bibitem{HOR} J. Horvath, \textit{Topological Vector Spaces and Distributions}, Addison-Wesley, Reading, Massachusetts, 1966.



\bibitem{BOHM1} A. Bohm, J. Math. Phys., {\bf 22} (1981) 2813-2823.



\bibitem{CG} O. Civitarese, M. Gadella, Phys. Rep., {\bf 396} (2004) 41-113.

\bibitem{CGO} E. Celeghini, M. Gadella, M.A. del Olmo, J. Math. Phys., {\bf 57} (2016) 072105.

\bibitem{CGO1} E. Celeghini, M. Gadella, M.A. del Olmo, J. Math. Phys., {\bf 59} (2018) 053502.

\bibitem{CGO2} E. Celeghini, M. Gadella, M.A. del Olmo, Acta Polytechnica (Prag), {\bf 57} (2017) 379-384.

\bibitem{GM} M. Gadella, R. de la Madrid, Int. J. Theor. Phys., {\bf 38} (1999) 93-113.

\bibitem{AGP} I.E. Antoniou, M. Gadella, G.P. Pronko, J. Math. Phys., {\bf 39} (1998) 2459-2475.

\bibitem{MH} A. Mondrag\'on, E. Hern\'andez, J. Phys. A: Math. Gen., {\bf 26} (1993) 5595-5611.

\bibitem{GL} M. Gadella, R. Laura, Int. J. Quant. Chem., {\bf 81} (2001) 307-320.

\bibitem{W} E. P. Wigner,\textit{ Group Theoretical Concepts and Methods in Elementary Particle Physics},
Gordon and Breach, New York, 1994, 37-38.

\bibitem{WI} E. P. Wigner, \textit{Symmetries and Reflections}, Indiana University Press, Bloomington, 1967,
38-39.



\bibitem{ALST} I. Antoniou, R. Laura, Z. Suchanecki, S. Tasaki, Physica A, {\bf 241} (1997) 737-772.


\bibitem{CAS} M. Castagnino, O. Lombardy, Philosophy of Science, {\bf 72} (2005) 764-776. 

\bibitem{CAS2} M. Castagnino, M. Gadella, Foundations of Physics, {\bf 36} (2006) 920-952.


\bibitem{CGIL} M. Castagnino, M. Gadella, R. Id Bet\'an, R. Laura, J. Phys. A: Math. Gen., {\bf 34} (2001) 10067-10083.

\bibitem{AGS} I. Antoniou, M. Gadella, Z. Suchanecki, {\it Some general properties of Liouville Spaces} in {\it Irreversibility and Causality}, Lecture Notes in Physics, {\bf 504}, Springer
Verlag, 1998, 38-56.

\bibitem{GA} M. Gadella, Foundations of Physics, {\bf 45} (2015) 177-197.




\bibitem{Kiefer} C. Kiefer, D. Polarski, Why do cosmological perturbations look classical to us?, Adv. Sci. Lett., \textbf{2} (2009) 164-173.

\bibitem{Reboiro}
R. Ram\'irez and M. Reboiro, Dynamics of finite dimensional non-hermitian systems with indefinite metric, J. Math. Phys., \textbf{60} (2019) 012106. 

\bibitem{Reboiro2}
R. Ram\'irez and M. Reboiro, Optimal spin squeezed steady state induced by the dynamics of non-hermtian Hamiltonians, 
Physica Scripta  (2019).


\end{thebibliography}
\end{document}